# Spin-torque oscillation in a magnetic insulator probed by a single-spin sensor

H. Zhang[1,2,†], M.J.H. Ku[1,2,†], F. Casola[1,2], C.H. Du[2], T. van der Sar[2,‡], M.C. Onbasli[3,4], C.A. Ross[3], Y. Tserkovnyak[5], A. Yacoby[2,6], R.L. Walsworth[1,2,7,*]

[1]Harvard-Smithsonian Center for Astrophysics, 60 Garden Street, Cambridge, MA 02138, USA.

[2]Department of Physics, Harvard University, 17 Oxford Street, Cambridge, MA 02138, USA.

[3]Department of Materials Science and Engineering, Massachusetts Institute of Technology, 77 Massachusetts Avenue, Cambridge, MA 02139, USA.

[4]Koç University, Department of Electrical and Electronics Engineering, Sarıyer, 34450 Istanbul, Turkey.

[5]Department of Physics and Astronomy, University of California, Los Angeles, 475 Portola Plaza, Los Angeles, CA 90095, USA.

[6]John A. Paulson School of Engineering and Applied Sciences, Harvard University, Cambridge, MA 02138, USA.

[7]Center for Brain Science, Harvard University, Cambridge, Massachusetts 02138, USA.

[†]These authors contributed equally to this work.

[‡]Present address: Kavli Institute of Nanoscience, Delft University of Technology, 2628CJ Delft, Netherlands.

*e-mail: rwalsworth@cfa.harvard.edu

**Coherent, self-sustained oscillation of magnetization in spin-torque oscillators (STOs)[1,2,3,4] is a promising source for on-chip, nanoscale generation of microwave magnetic fields. Such fields could be used for local excitation of spin-wave resonances, control of spin qubits, and studies of paramagnetic resonance. However, local characterization of fields emitted by an STO has remained an outstanding challenge. Here, we use the spin of a single nitrogen-vacancy (NV) defect in diamond to probe the magnetic fields generated by an STO in a microbar of ferromagnetic insulator yttrium-iron-garnet (YIG). The combined spectral resolution and sensitivity of the NV sensor allows us to resolve multiple spin-wave modes and characterize their damping. When damping is decreased sufficiently via spin injection, the modes auto-oscillate, as indicated by a strongly reduced linewidth, a diverging magnetic power spectral density, and synchronization of the STO frequency to an external microwave source. These results open the way for quantitative, nanoscale mapping of the microwave signals generated by STOs, as well as harnessing STOs as local probes of mesoscopic spin systems.**

Spin-torque oscillators (STOs) have been proposed as on-chip sources of spin waves[2], as nanoscale microwave generators[2,3,4], and as building blocks in neural networks[2,3,5]. While optical methods[6,7,8] such as Brillouin light scattering and magneto-optical Kerr effect microscopy have been used to spatially investigate the magnetization dynamics of STOs, the detection of on-chip STO microwave fields with a probe that allows nanoscale spatial imaging and MHz spectral resolution has remained an outstanding challenge. Recently, the electron spin associated with the nitrogen-vacancy (NV) defect in diamond has emerged as a sensitive magnetic-field sensor[9] that allows nanometer-scale spatial resolution[10,11], sub-Hz spectral resolution[12], and excellent magnetic-field sensitivity[13]. Here, we use NV magnetometry to study the local magnetic fields generated by an STO in a YIG microbar that can be driven into auto-oscillation by a spin-current injected via the spin-Hall effect in a platinum thin film[8,14,15].

To locally detect the STO magnetic fields, we position a diamond nanobeam containing an individually addressable NV sensor at ~100 nm from a Pt/YIG hybrid microstructure (Fig. 1a). Au electrical leads supply the DC current $I_{dc}$ to the Pt wire for spin injection. A nearby Au stripline (Fig. 1b) delivers microwave signals for both control of the NV spin state and for microwave-driving of spin-wave modes in the YIG bar. Further information on NV physics relevant to sensing of magnetic fields can be found in refs. 9-13,16-18, and details of the fabrication processes can be found in the Methods and in Supplementary Information 1.

We start by probing the spin-wave spectrum of the YIG micro-magnet using microwave excitation. We sweep the frequency of a microwave drive field and use the NV spin sensor to detect changes in the stray static magnetic field due to changes in the YIG magnetization upon exciting a spin-wave resonance[16]. Fig. 1c depicts the sensing sequence, and Fig. 1d shows the spectrum at zero DC current. We observe multiple spectral peaks, the most prominent of which persists as the strongest mode throughout the entire sweep range of the external magnetic field $B_{ext}$. We attribute this peak to the spatially homogeneous (n=1) ferromagnetic resonance (FMR) of the YIG bar as it couples efficiently to our microwave drive field. The centre frequency of this mode vs $B_{ext}$ follows a Kittel-like dependence (Fig. 1e).

Next, we demonstrate control of spin-wave resonance damping by injecting a spin current into the YIG via the Pt contact. As illustrated in Fig. 2a, the effect of a spin current on the dynamics of the magnetization can be described as a spin-orbit torque acting on the YIG magnetization[19]. Depending on the relative orientation between the injected spins and the equilibrium magnetization, the spin-orbit torque can either reduce or enhance the effective damping of precessional motion of the magnetization vector (see Supplementary Information 3). We observe such damping modification by measuring the response of the YIG stray field to the microwave drive as a function of $I_{dc}$. In Fig. 2b, we show example FMR spectra of the YIG stray field $\Delta B_{||}$, normalized by the microwave drive amplitude $b_1$, as a function of the drive frequency at several $I_{dc}$. A strong dependence of the peak amplitude on $I_{dc}$ indicates that damping is reduced (enhanced) for positive (negative) $I_{dc}$. In addition, we observe a shift in the resonance frequency as a function of $I_{dc}$ (Fig. 2b) that is well described by a second-order polynomial[8]. We attribute the quadratic (symmetric) part of this polynomial to Joule heating, while the linear part may be attributed to a combination of effects such as an Oersted field generated by the current in the Pt, and a change in the YIG magnetization caused by the spin-orbit torque (see Supplementary Information 4).

When the intrinsic magnetic damping is compensated by the anti-damping torque exerted by the injected spin current, we expect an increase in the rate of change of the FMR peak amplitude as a function of the microwave drive power. Figure 2c plots the on-resonance peak amplitude of $\Delta B_{||}$, extracted from data such as those in Fig. 2b, as a function of the microwave drive power $b_1^2$ at different $I_{dc}$. We observe two distinct regimes: for $I_{dc} \lesssim 4$ mA, the peak amplitude increases approximately linearly over a large range of microwave drive power, while for $I_{dc} \gtrsim 4$ mA, the signal increases sharply and then saturates. The distinction between these two regimes is more evident in Fig. 2d, which plots the inverse of the initial slopes of $\Delta B_{||}$ vs $b_1^2$ (i.e., $db_1^2 / d\Delta B_{||}$) as a function of $I_{dc}$. The trend of the inverse on-resonance slopes quantifies the evolution of the effective damping coefficient $\alpha_{eff}$, which monotonically changes with $(1/\chi'')^2 \propto (b_1/m)^2 \sim db_1^2/d\Delta B_{||}$, where $\chi''$ is the imaginary part of the YIG bar's magnetic susceptibility, and $m$ is the oscillation amplitude of the transverse magnetization[20]. In Fig. 2d we observe

$db_1^2/d\Delta B_\parallel$ decreases as a function of $I_{dc}$ (diagonal dashed line) until a certain threshold, after which it plateaus (horizontal dashed line). We interpret the plateau to be due to auto-oscillation of a spin-torque oscillator (STO), with the crossing of the dashed lines corresponding to the STO onset threshold current ~ 3.8 mA. Note that this threshold current value agrees well with an independent estimate obtained from Fig. 4a.

To study auto-oscillation of the spin-wave modes in the YIG bar, we characterize the power spectral density of the magnetic-noise generated by the modes as a function of $I_{dc}$ and in the absence of microwave excitation. We use the NV spin as a field-tunable spectrometer via a technique known as NV spin relaxometry[16], where the NV spin relaxation rates $\Gamma$ are measured to quantify the magnetic-noise power spectral density $B^2(\omega)$ at the NV $m_s = 0 \leftrightarrow \pm 1$ transition frequencies $\omega_\pm = 2\pi(D_{gs} \pm \gamma B_{ext})$ via the relation $\Gamma_\pm = \frac{\gamma^2}{2}B^2(\omega_\pm)$. Here, $\gamma$=2.8 MHz/G is the NV gyromagnetic ratio, $D_{gs}$ =2.87 GHz is the NV zero-field splitting, and $B_{ext}$ is the external static magnetic field aligned with the NV axis. To characterize NV spin relaxation, we prepare the NV spin in the $m_S$=0 state and determine the spin-relaxation rate $\Gamma$ in the presence of YIG bar magnetic noise by measuring the spin-dependent photoluminescence $PL(\tau, \omega) = PL(\tau = 0)e^{-\Gamma(\omega)\tau}$ after a hold time $\tau$ (Fig. 3a). In Fig. 3b, we show an example NV spin relaxometry experiment at $I_{dc}$=5.8 mA, where we measure the PL at a fixed $\tau$ as we vary the external magnetic field that sweeps the lower transition frequency $\omega_-$ over several spin-wave resonances. In this situation, the rate $\Gamma \approx \Gamma_-$; $\Gamma_+$ is negligible as the density of thermal magnons is suppressed at energy $\omega_+$ which is far detuned from spin-wave resonances. The magnetic field fluctuations produced by the different YIG spin-wave modes increase the NV spin relaxation rate, which results in decreased NV PL[21,22]. By performing this measurement at multiple $I_{dc}$, we map the noise spectrum of spin-waves as a function of $B_{ext}$ and $I_{dc}$ (Fig. 3c). The red/blue stars denote the locations of prominent spin-wave resonances (obtained by fitting the peak centres at each $I_{dc}$), which we call $STO_1$ and $STO_2$. We identify $STO_1$ as the spatially homogeneous (n=1) FMR and $STO_2$ as a higher order (n=2) spin wave mode (Supplementary Information 5). The spectral resolution of the NV sensor allows us to zoom in closely on the regions where the two spin-waves approach each other (Figs. 3d and 3e), and observe what seems to be a mode anti-crossing, hinting at hybridization of the spin-wave modes due to mode interactions. Micromagnetic simulations elucidate the nature of the modes and point to the possibility of mode mixing (Supplementary Information 5). We estimate a mode coupling strength of about 10 MHz (Supplementary Information 6), which is larger than the linewidth of the individual modes (see Fig. 4b).

To quantitatively study the power spectral density of the magnetic field noise generated by $STO_1$ and $STO_2$, we tune the NV transition frequency into resonance with the modes by adjusting $B_{ext}$ and extract the NV spin relaxation rate $\Gamma$ as a function of $I_{dc}$ (Fig. 4a). As we change $I_{dc}$, we observe a dramatic increase in the STO magnetic-noise power spectral density of up to three orders of magnitude, a key signature of auto-oscillation[23]. The inset of Fig. 4a maps $1/\Gamma$ as a function of $I_{dc}$, and we use the intersection of a linear fit of $1/\Gamma$ at low current with $1/\Gamma = 0$ to indicate the onset (threshold) of auto-oscillation, following the relation[24,25] $1/\Gamma \propto 1/p \propto 1 - I_{dc}/I_{th}$, where $p$ is the peak power spectral density emitted by the STO and $I_{th}$ is the threshold current. For $STO_1$, we estimate $I_{th1} \approx 3.5$ mA, close to the estimate made above from the stray-field magnetometry measurements in Fig. 3d. For $STO_2$, we obtain a higher threshold current $I_{th2} \approx 4.4$ mA. A strong correlation between linewidth reduction (Fig. 4b) and divergence of magnetic fluctuations (Fig. 4a) is consistent with Landau-Lifshitz-Gilbert phenomenology: that is, the spin-orbit torque reduces the damping torque and the associated STO linewidth. As we increase $I_{dc}$ further, we surprisingly observe a reduction in the power spectral density accompanied by linewidth broadening and the appearance of a higher-order STO (STO* as shown in Fig.

3b). This may imply that strong spin injection introduces an additional magnon decay channel and the magnetic system is approaching a re-thermalization scenario[26], though we leave a study of this phenomenon to future work.

Finally, we demonstrate that the Pt/YIG STO can be synchronized with an external microwave source, as observed previously for STOs in metallic ferromagnets[26]. We use the measurement scheme shown in Fig. 5a, which is an NV spin relaxometry measurement with an added microwave drive field. We sweep the frequency $f_{MW}$ of this drive field around the free-running STO frequency. By monitoring the magnetic-noise power spectral density at the NV transition frequency, we observe locking of the STO over a frequency interval $\Delta f_s$ (Fig. 5b). Figures 5c and 5d show that the locking interval increases approximately linearly with the drive amplitude $b_1$ as expected for frequency-locked oscillators. We observe an increase of the synchronization bandwidth for larger $I_{dc}$ (Fig. 5d). Frequency locking to an external microwave source can be used to quickly tune the STO in and out of resonance with the control frequency of a target system.

In summary, we used the spin of a single NV center in diamond as a nanoscale magnetic sensor to measure the local magnetic fields generated by spin-torque oscillators (STOs) driven by spin-current in a Pt/YIG hybrid microstructure. We demonstrated STO auto-oscillation[15] in this magnetic insulator using three independent methods: suppression of the effective damping torque, divergence of the power spectral density at the STO frequency, and STO synchronization to an external microwave source. High spectral resolution is a key capability of NV sensing and can be further improved below ~1 Hz by, for example, quantum interpolation[27], synchronized readout[12], or modest cooling[17]. Thus in future work, NV sensors should be able to provide access to the sub-Hz regime of advanced research on STOs[3], as well as nanoscale spatial characterization of STO-generated magnetic fields[10,11]. Spatial mapping at such length scales would provide access to locations of large STO magnetic-field intensity, with the potential to use an STO to drive magnetic excitations in other systems of interest, such as spin waveguides[28] and spin qubits[29]. Finally, studies of spin-torque oscillation may provide insight into phenomena such as magnon thermodynamics[30], strongly-correlated many body physics[18], and control over magnetic phase transitions[31].

## Methods

**Hybrid Pt/YIG device preparations**. Fabrication of the Pt/YIG device starts with a 17-nm YIG film epitaxially grown on a (111) orientation GGG substrate using pulsed laser deposition[32]. A 10 nm layer of Platinum (Pt) is sputtered on top of the YIG film, which is first cleaned by an Ar+ plasma at a pressure below $5\times10^{-8}$ Torr to ensure good Pt purity. The Pt/YIG stripe is defined by electron-beam lithography (Elionix F125, 125 kV) with a PMMA (495A2, ~ 30 nm)/HSQ (XR-1541-006, ~250 nm and FOX-16, ~ 500 nm) resist stack, followed by developing in 25% TMAH. Ar+ ion milling is used to transfer the pattern onto the substrate and form the Pt/YIG hybrid microstructure. Finally, leads for DC current and microwave driving are defined by electron-beam lithography and e-beam evaporation techniques. See Supplementary Information 1 for further details.

**Experimental setup and nanobeam fabrication.** The experimental setup is based on a home-built laser scanning confocal microscope, which has been described previously[11]. As part of the experimental sensing platform, we pattern bulk diamond containing NVs into a nanobeam structure[33] and place it close to the sample of interest, with a single NV sensor within about 100 nm of the Pt/YIG microstructure to access the sample's relatively weak localized fields.

# References

1. Chumak, A. V., Vasyuchka, V. I., Serga, A. A. & Hillebrands, B. Magnon spintronics. *Nat. Phys.* **11,** 453–461 (2015).

2. Locatelli, N., Cros, V. & Grollier, J. Spin-torque building blocks. *Nat. Mater.* **13,** 11–20 (2014).

3. Chen, T. *et al.* Spin-Torque and Spin-Hall Nano-Oscillators. *Proc. IEEE* **104,** 1919–1945 (2016).

4. Demidov, V. E. *et al.* Magnetization oscillations and waves driven by pure spin currents. *Phys. Rep.* **673**, 1-31 (2017).

5. Torrejon, J. *et al*. Neuromorphic computing with nanoscale spintronic oscillators. *Nature* **547**, 428–431 (2017)

6. Demokritov, S. O. *et al.* Bose-Einstein condensation of quasi-equilibrium magnons at room temperature under pumping. *Nature* **443,** 430–433 (2006).

7. Montazeri, M. *et al.* Magneto-optical investigation of spin–orbit torques in metallic and insulating magnetic heterostructures. *Nat. Commun.* **6,** 8958 (2015).

8. Demidov, V. E. *et al.* Direct observation of dynamic modes excited in a magnetic insulator by pure spin current. *Sci. Rep.* **6,** 32781 (2016).

9. Maze, J. R. *et al.* Nanoscale magnetic sensing with an individual electronic spin in diamond. *Nature* **455,** 644–647 (2008).

10. Grinolds, M. S. *et al.* Subnanometre resolution in three-dimensional magnetic resonance imaging of individual dark spins. *Nat. Nanotech.* **9,** 279–284 (2014).

11. Arai, K. *et al.* Fourier magnetic imaging with nanoscale resolution and compressed sensing speed-up using electronic spins in diamond. *Nat. Nanotech.* **10,** 859-864 (2015).

12. Glenn, D. R. *et al.* High Resolution Magnetic Resonance Spectroscopy Using Solid-State Spins. *Nature* **555,** 351 (2018).

13. Rondin, L. *et al.* Magnetometry with nitrogen-vacancy defects in diamond. *Rep. Prog. Phys.* **77,** 056503 (2014).

14. Hamadeh, A. *et al*. Full Control of the Spin-Wave Damping in a Magnetic Insulator Using Spin-Orbit Torque. Phys. Rev. Lett. **113**, 197203 (2014).

15. Collet, M. *et al.* Generation of coherent spin-wave modes in yttrium iron garnet microdiscs by spin-orbit torque. *Nat. Commun.* **7,** 10377 (2016).

16. van der Sar, T., Casola, F., Walsworth, R. & Yacoby, A. Nanometre-scale probing of spin waves using single-electron spins. *Nat. Commun.* **6,** 7886 (2015).

17. Bar-Gill, N. *et al.* Solid-state electronic spin coherence time approaching one second. *Nat. Commun.* **4,** 1743 (2013).

18. Casola, F., van der Sar, T., & Yacoby, A. Probing condensed matter physics with magnetometry based on nitrogen-vacancy centres in diamond. *Nature Review Materials* **3,** 17088 (2018).

19. Slonczewski, J. C. Current-driven excitation of magnetic multilayers. *J. Magn. Magn. Mater.* **159,** L1–L7 (1996).

20. Bailleul, M., Höllinger, R., & Fermon, C. Microwave spectrum of square Permalloy dots:

Quasisaturated state. *Phys. Rev. B* **73**, 104424 (2006).

21. Wolfe, C. S. *et al.* Off-resonant manipulation of spins in diamond via precessing magnetization of a proximal ferromagnet. *Phys. Rev. B* **89**, 180406(R) (2014).

22. Du, C. *et al.* Control and local measurement of the spin chemical potential in a magnetic insulator. *Science* **357,** 195 (2017).

23. Demidov, V. E. *et al.* Magnetic nano-oscillator driven by pure spin current. *Nat. Mater.* **11,** 1028–1031 (2012).

24. Slavin, A. & Tiberkevich, V. Nonlinear auto-oscillator theory of microwave generation by spin-polarized current. *IEEE Trans. Magn.* **45,** 1875–1918 (2009).

25. Hamadeh, A. *et al.* Autonomous and forced dynamics in a spin-transfer nano-oscillator: Quantitative magnetic-resonance force microscopy. *Phys. Rev. B* **85,** 140408 (2012).

26. Demidov, V. E. *et al.* Synchronization of spin Hall nano-oscillators to external microwave signals. *Nat. Commun.* **5,** 3179 (2014).

27. Ajoy, A. *et al.* Quantum Interpolation for High Resolution Sensing. *Proc. Natl. Acad. Sci. USA* **114**, 2149-2153 (2017).

28. Collet, M. *et al*. Spin-wave propagation in ultra-thin YIG based waveguides. *Appl. Phys. Lett.* **110**, 092408 (2017).

29. Sutton, B. & Datta, S. Manipulating quantum information with spin torque. *Sci. Reports* **5,** 17912 (2015).

30. Safranski, C. *et al.* Spin caloritronic nano-oscillator. *Nat. Commun.* **8,** 117 (2017).

31. Giamarchi, T., Rüegg, C. & Tchernyshyov, O. Bose-Einstein Condensation in Magnetic Insulators. *Nat. Phys.* **4,** 198 (2008).

32. Lang, M. e*t al.* Proximity Induced High-Temperature Magnetic Order in Topological Insulator - Ferrimagnetic Insulator Heterostructure. *Nano Lett*. **14**, 3459 (2014).

33. Burek, M. J. *et al*. Free-standing mechanical and photonic nanostructures in single-crystal diamond. *Nano Lett*. **12**, 6084 (2012).

## Acknowledgements

The authors acknowledge the provision of diamond samples by Element 6, assistance with nanobeam fabrication from M. Warner and M. Burek, the use of a setup for nanobeam transfer from P. Kim, use of the ion mill facility in the J. Moodera lab, and experimental assistance from K. Arai, M. Han, and J.-C. Jaskula. This material is based upon work supported by, or in part by, the United States Army Research Laboratory and the United States Army Research Office under Contract/Grants No. W911NF1510548 and No. W911NF1110400. A.Y. acknowledges support from the Army Research Office under Grant Number W911NF-17-1-0023. The views and conclusions contained in this document are those of the authors and should not be interpreted as representing the official policies, either expressed or implied, of the Army Research Office or the U.S. Government. The U.S. Government is authorized to reproduce and distribute reprints for Government purposes notwithstanding any copyright notation herein. Work at the Massachusetts Institute of Technology was supported by the Solid-State Solar-Thermal Energy Conversion Center (S3TEC), an Energy Frontier Research Center funded by DOE, Office of Science, BES under award no. DE-SC0001299/DE-FG02-09ER46577. Work at the University of California, Los Angeles, is supported by the U.S. Department of Energy (DOE), Office of Basic Energy Sciences (BES) under award no. DE-SC0012190. F.C. acknowledges support from the Swiss National Science Foundation grant no. P300P2-158417. This research is also funded in part by the Gordon and Betty Moore Foundation's EPiQS Initiative through Grant GBMF4531, the STC Center for Integrated Quantum Materials, NSF Grant No. DMR-1231319, and by the National Science Foundation under Grant No. EFMA-1542807. This work was performed in part at the Center for Nanoscale Systems (CNS), a member of the National Nanotechnology Coordinated Infrastructure Network (NNCI), which is supported by the National Science Foundation under NSF award no. 1541959. CNS is part of Harvard University.

## Author contributions

H.Z. and M.J.H.K. contributed equally to this work. H.Z, M.J.H.K, A.Y., and R.L.W. conceived the project. R.L.W. and A.Y. supervised the project. H.Z. and M.J.H.K developed measurement protocols, built the experimental setup for NV measurement, performed the measurements, and analyzed the data. H.Z. fabricated the Pt/YIG device and performed micromagnetic simulations. F.C. and H.Z. developed the nanobeam platform and fabricated the nanobeams. F. C. helped with micromagnetic simulations. M.C.O. and C.A.R. provided the YIG sample. H.Z., M.J.H.K., T.v.d.S., F.C., C.H.D., Y.T., A.Y., and R.L.W. contributed to the interpretation. H.Z., M.J.H.K., T.v.d.S., C.H.D., and R.L.W. wrote the manuscript with the help from all co-authors.

## Competing interests

The authors declare no competing financial interests

## Materials & Correspondence.

Correspondence to R. L. Walsworth (rwalsworth@cfa.harvard.edu).

# Figures

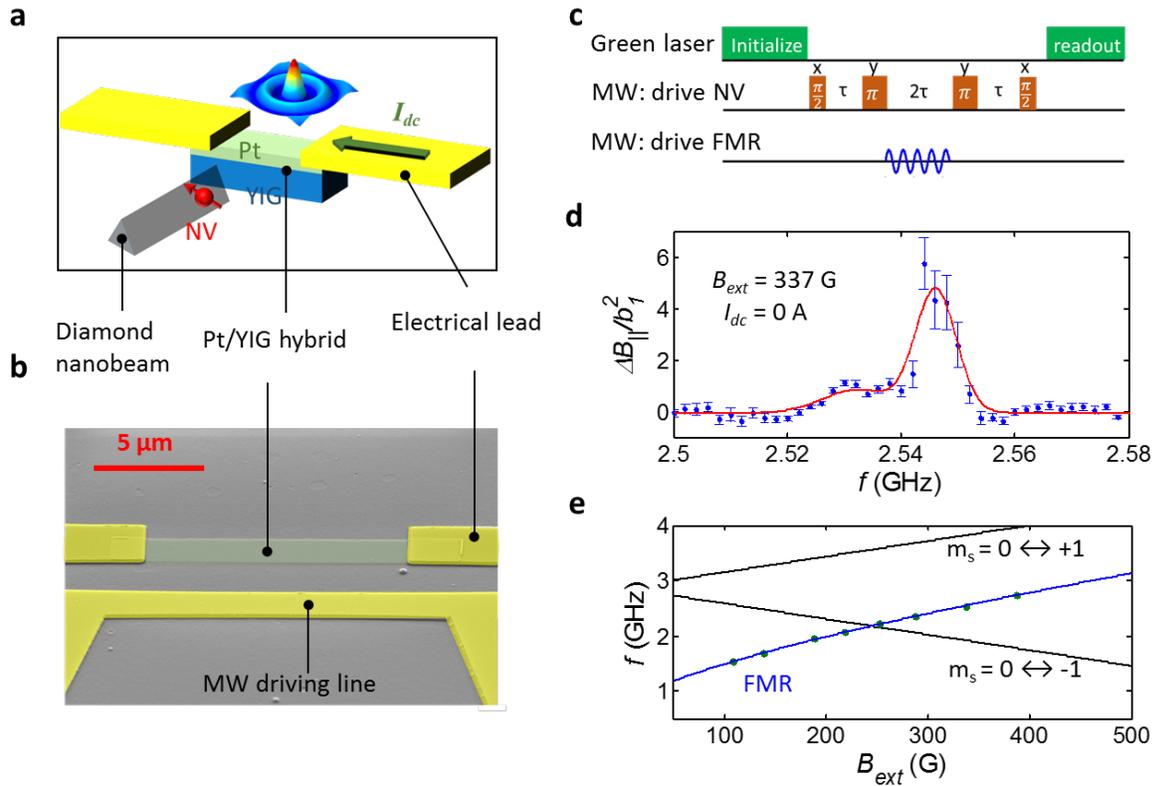

**Figure 1 | Local probing of a Pt/YIG spin-torque oscillator using a single nitrogen-vacancy (NV) sensor spin in diamond. a,** Schematic illustration of the device: a diamond nanobeam containing a single NV spin is positioned ~100 nm from a hybrid Pt/YIG structure (2.5×9 µm$^2$). The Pt (YIG) thickness is 10 nm (17 nm). Au wires provide electrical contact to the Pt film. **b,** False-colored scanning electron micrograph of the device (before positioning the diamond nanobeam). **c,** Measurement sequence for stray-field magnetometry of spin-wave resonances. A green laser pulse initializes the NV spin. The first microwave π/2 pulse prepares an NV spin superposition, followed by a spin-echo sequence with two π pulses. A spin-wave (FMR) drive is applied during the central 2τ period. A change in the YIG stray field $\Delta B_{\parallel}$ imparts a phase $\phi = \gamma \Delta B_{\parallel} 2\tau$ on the NV spin state, where γ=2.8 MHz/G is the NV gyromagnetic ratio. The final π/2 pulse converts this phase to an NV spin population difference, which is read out via spin-dependent photoluminescence. A free precession time τ ≈ 5.5 µs is used for such stray-field magnetometry. **d,** Example of YIG spin-wave resonances measured with the pulse sequence in **c**, at applied static magnetic field $B_{ext}$ = 337 G aligned with the NV axis. Plotted is the NV-measured stray static magnetic field along the NV axis, $\Delta B_{\parallel}$, as a function of the spin-wave drive frequency. The signal is normalized by $b_1^2$, which is proportional the spin-wave drive power. $b_1$ is independently measured on-chip using the same NV sensor via tuning $B_{ext}$ to bring the $m_s = 0 \leftrightarrow -1$ transition on resonance with the drive field and measuring the NV Rabi frequency[16]. Blue dots: data. Red line: double Gaussian fit, yielding FWHM = 8.5(6) MHz for the dominant mode attributed to the spatially homogeneous (n=1) ferromagnetic resonance (FMR) of the YIG bar. **e,** Green dots: Magnetic-field ($B_{ext}$) dependence of the fundamental spin-wave resonance frequency extracted from fits to measurements such as shown in **d**. Blue line: fit reveals characteristic Kittel-like behavior of FMR. Black lines: NV transition frequencies corresponding to the $m_s = 0 \leftrightarrow \pm 1$ transitions. NV-spin manipulation pulses are applied on the $m_s = 0 \leftrightarrow +1$ transition.

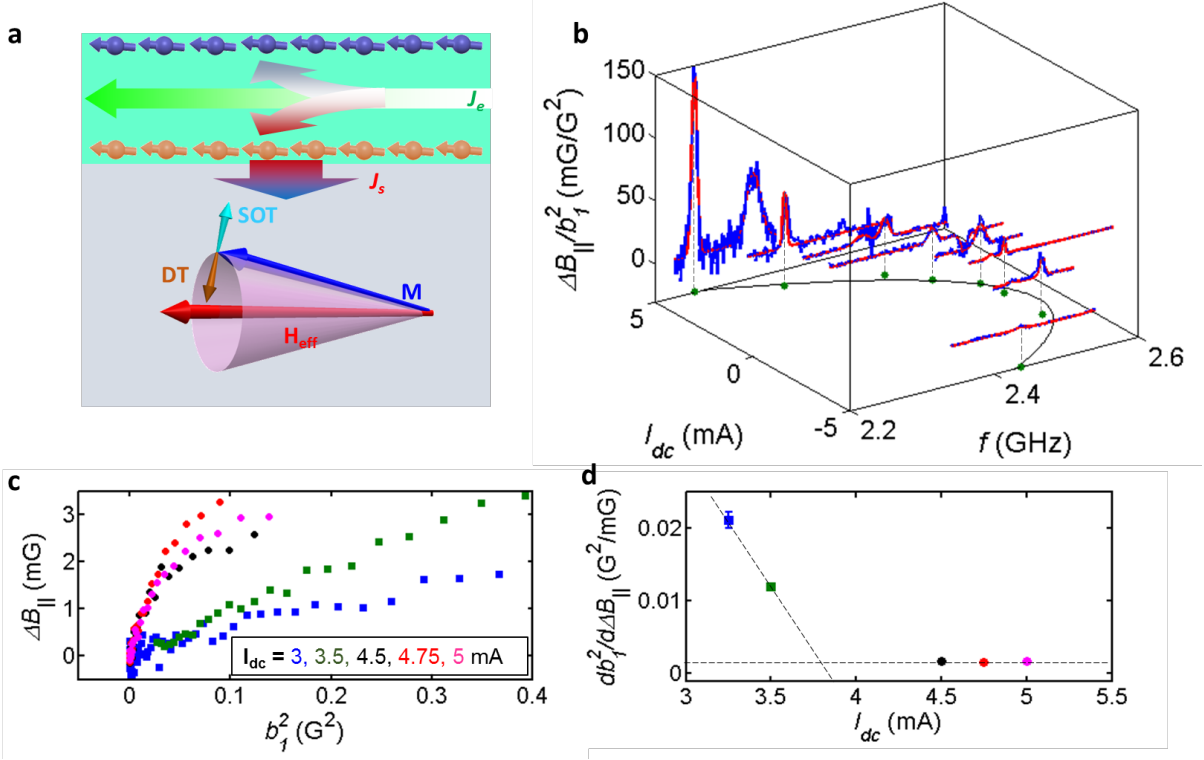

**Figure 2 | Controlling spin-wave damping using electrically controlled spin injection. a**, Sketch of the magnetization dynamics of the Pt/YIG device under the influence of a spin current. The electrical current ($J_e$) injects a spin current ($J_s$) into the YIG, leading to a spin-orbit torque (labeled **SOT**) that either reduces or enhances magnetic damping depending on the relative orientation between the injected spins and the magnetization **M**. **DT** denotes the (intrinsic) damping torque. **b**, NV-measured, microwave-driven spin-wave resonance spectra in the YIG as a function of the DC current $I_{dc}$ through the Pt. (Measurement sequence shown in Fig. 1c.) Blue traces: normalized change in YIG stray field $\Delta B_{||}/b_1^2$ as a function of microwave driving frequency for $I_{dc}$ = 5, 4, 3, 2, 1, 0, -2, -5 mA. Red lines: double Gaussian fit to data. Green dots: centre frequency of the fundamental spin-wave mode vs. $I_{dc}$. Black curve: parabolic fit to green dots. **c**, On-resonance $\Delta B_{||}$ as a function of microwave driving power $b_1^2$ for different values of $I_{dc}$. Black, red, and pink dots correspond to $I_{dc}$ = 4.5, 4.75, and 5 mA, for which the initial slopes ($d\Delta B_{||}/db_1^2$) have no discernable difference. Blue and green squares correspond to $I_{dc}$ = 3 and 3.5 mA, for which initial slopes are significantly smaller. **d**, Plot of the inverse of the initial slopes, i.e., $db_1^2/d\Delta B_{||}$, as a function of $I_{dc}$. Diagonal and horizontal dashed lines serve as eye guide to illustrate that there exists a current threshold as onset of an auto-oscillating spin torque oscillator (STO).

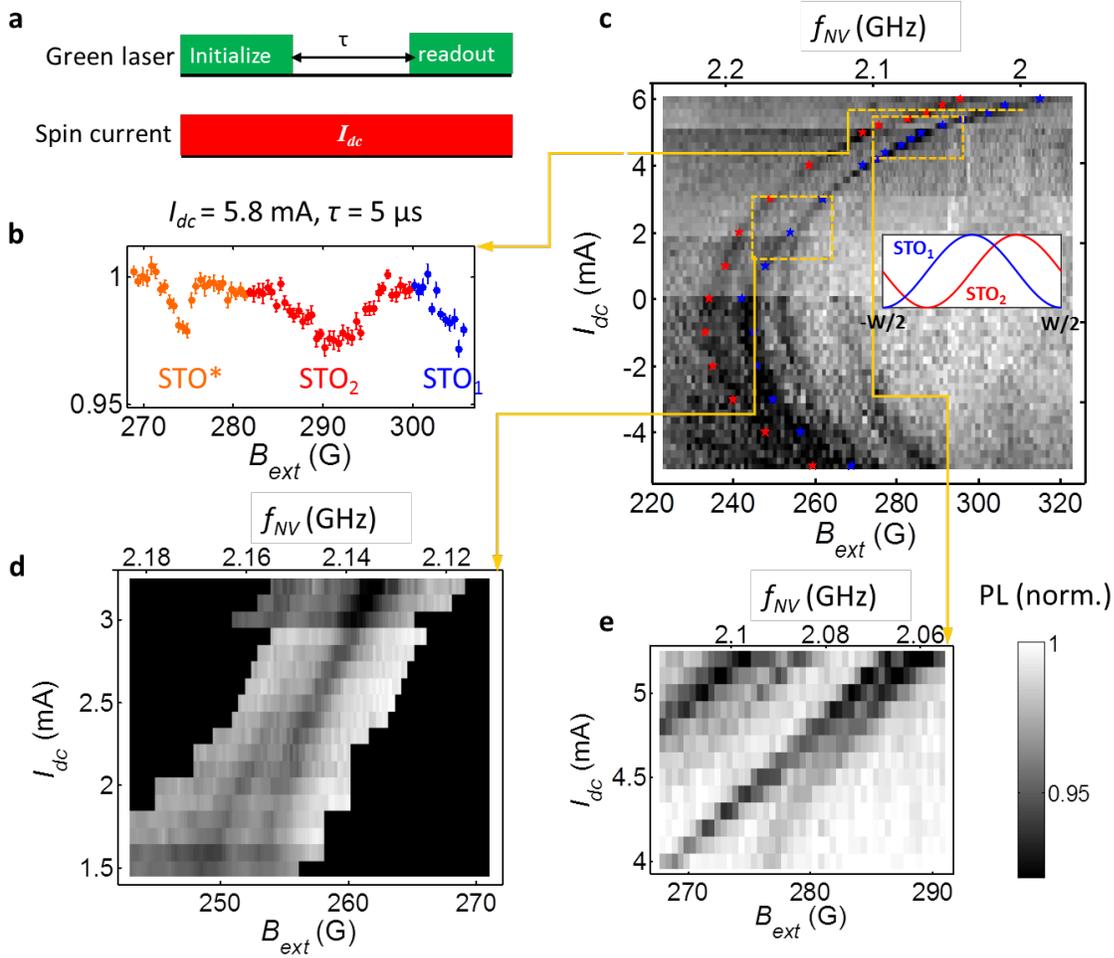

**Figure 3 | Spin-wave noise spectroscopy via NV spin relaxometry**. **a**, NV spin relaxometry measurement sequence. The NV spin is initialized into $m_s = 0$ by a green laser pulse and let to relax for a time $\tau$, after which the spin population is characterized via the spin-dependent PL during a laser readout pulse. The current $I_{dc}$ enhances or reduces spin-wave damping in the YIG bar and alters the power spectral density of the magnetic-field noise accordingly. Noise that is resonant with an NV transition frequency causes NV spin relaxation. **b**, Example NV spin relaxometry measurement at $I_{dc}$=5.8 mA and $\tau$=5 μs. By tuning the magnetic field $B_{ext}$, the frequency of the $m_s = 0 \leftrightarrow -1$ transition is swept over three spin-wave (SW) modes in the YIG, whose field-noise causes strong NV spin relaxation and thus dips in the normalized PL signal. **c**, Performing the measurement shown in panel **b** for different $I_{dc}$ yields a 2D plot of PL vs $I_{dc}$ and $B_{ext}$ that displays the presence and dispersion of spin-torque oscillators (STOs) in the system. Different delay times $\tau$ of 150 μs, 50 μs, 15 μs, 5 μs, and 3 μs are used for the different $I_{dc}$ ranges of [-5 mA:0 mA], [0.2 mA:1.8 mA], [2 mA:3 mA], [3.2 mA:5 mA], and [5.2 mA:6 mA], respectively. Top horizontal axis shows the $m_s = 0 \leftrightarrow -1$ transition frequency at corresponding $B_{ext}$. Blue stars indicate fits of peak centres for the first resonance on the left-hand-side (STO$_1$), while red stars are fits of peak centres for the second (STO$_2$). These two STOs are also indicated in panel **b**. Note that an additional oscillator (data points are orange in color and designated as STO* in panel **b**) appears when $I_{dc}$ = 5.8 mA and persists for higher current. Inset illustrates mode spatial distribution of STO$_1$ and STO$_2$ along width of Pt/YIG microstructure (W). **d&e**. Zoomed-in, high-resolution views of **c**, where spin-wave modes are observed to approach each other.

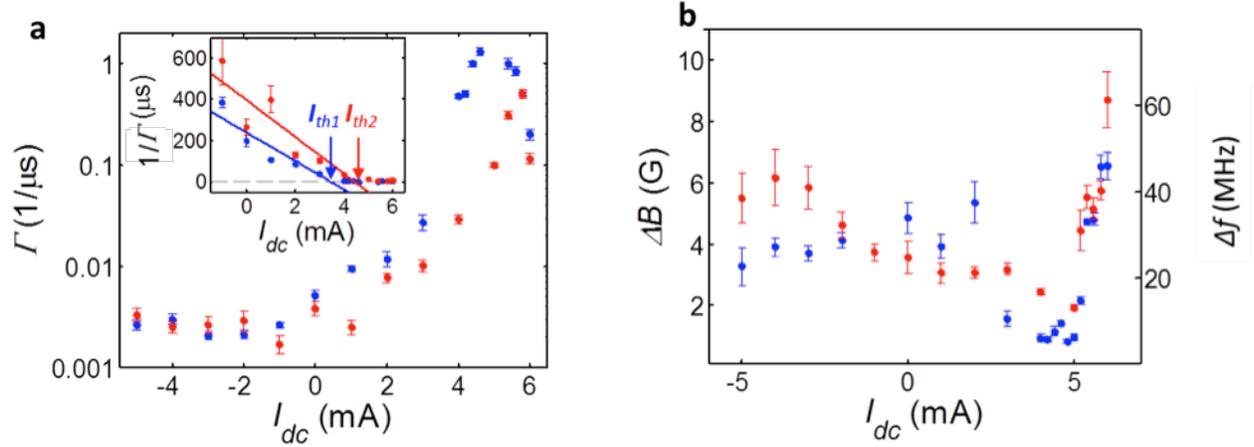

**Figure 4 | Detection of spin-torque auto-oscillation by NV spin relaxometry**. **a**, NV spin relaxation rate ($\Gamma$) is measured at the current and magnetic field values indicated by the blue and red stars in Fig. 3c, where the spin-torque oscillators (STOs) are resonant with the $m_s = 0 \leftrightarrow -1$ transition frequency. At each of these current and magnetic field values, we sweep $\tau$, perform an NV spin relaxometry measurement sequence (Fig. 3a), and extract the exponential decay time constant $\Gamma$. $\Gamma_{STO1}$ (red) and $\Gamma_{STO2}$ (blue) are plotted as a function of $I_{dc}$. The dramatic order-of-magnitude increase of the relaxation rate above $I_{dc} \sim 3$ mA indicates spin-torque induced auto-oscillation of the STOs. Inset shows $1/\Gamma$ vs $I_{dc}$ for both STOs. Linear fits at low current ($I_{dc} < 4$ mA) intersect with $T_1 = 0$ at $I_{th1} = 3.5$ mA and $I_{th2} = 4.4$ mA, which we define as the auto-oscillation threshold currents. **b**, Measured STO linewidth $\Delta B$ as a function of $I_{dc}$ for $STO_1$ (blue dots) and $STO_2$ (red dots). The vertical axis on the right gives the linewidth in frequency (MHz), calculated from $\Delta B$ using the Kittel relation at $B_{ext} \sim 250$ G and $I_{dc} = 0$.

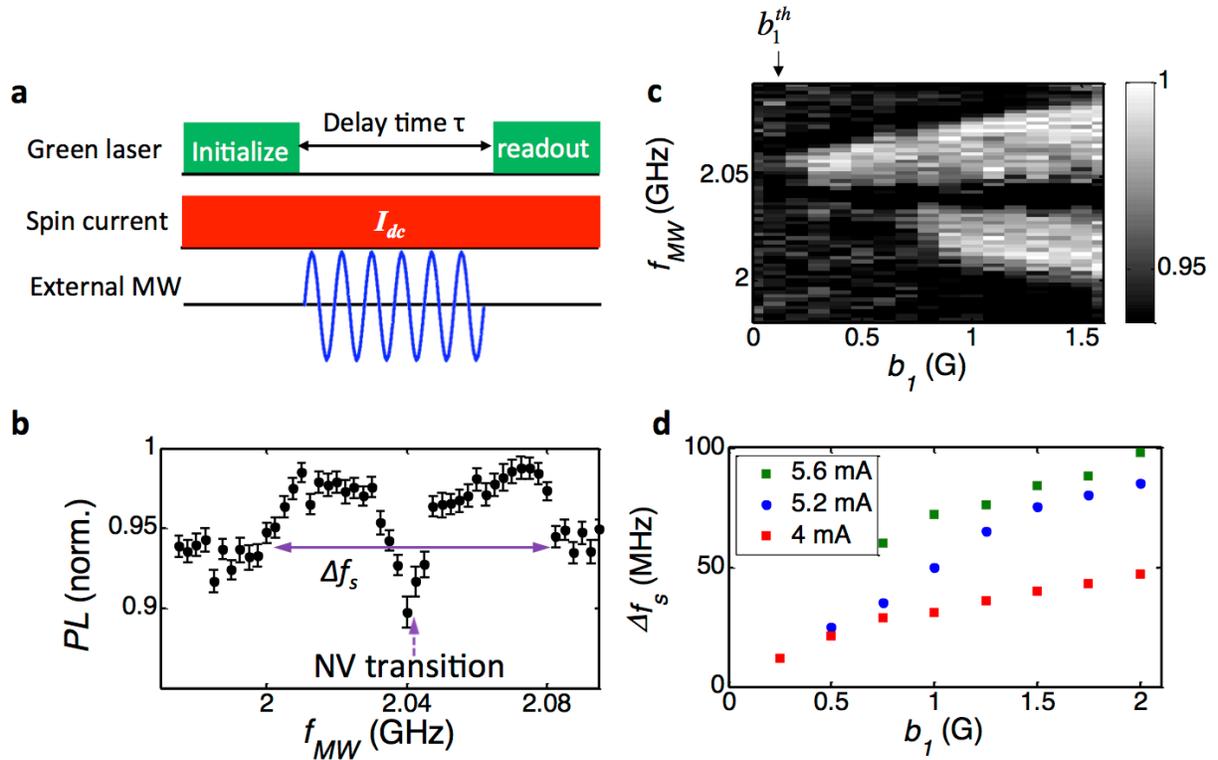

**Figure 5 | Locking STO frequency to an external microwave (MW) source**. **a**, NV spin relaxometry measurement sequence as in Fig. 3a, with added MW drive. For synchronization measurement, $B_{ext}$ is tuned such that the NV $m_s = 0 \leftrightarrow -1$ transition coincides with STO resonance. **b**, Measured NV photoluminescence (PL) as a function of the MW drive frequency $f_{MW}$, at $I_{dc} = 5.2$ mA, $B_{ext} = 292$ G, and MW drive amplitude $b_1 = 1.5$ G. When the MW drive is resonant with the NV transition frequency, a dip in the PL is observed because the driving depletes the $m_s=0$ population. Over a frequency interval $\Delta f_s$ the STO can be locked to the MW drive and thus detuned from the NV transition, thereby decreasing the NV spin relaxation and correspondingly increasing the measured PL. When the MW drive frequency is detuned beyond the locking interval (i.e., synchronization bandwidth), the STO remains resonant with the NV transition frequency, leading to strong NV-spin relaxation and a corresponding reduced PL. (See Supplementary Information 8 for detailed data analysis.) **c**, 2D map of PL vs $f_{MW}$ and MW drive amplitude $b_1$. The synchronization bandwidth increases linearly with $b_1$. **d**, Synchronization bandwidth vs $b_1$ at different $I_{dc}$ (4, 5.2, and 5.6 mA).